\documentclass[pra,twocolumn,preprintnumbers]{revtex4} 

\usepackage{graphicx}
\usepackage{bm}
\usepackage{amsmath}
\usepackage{amssymb}
\usepackage{amsfonts}
\usepackage{fancyhdr}
\usepackage{slashed}
\usepackage{latexsym,epsfig,bbm}

\newcommand{\bra}[1]{\langle{#1}|}
\newcommand{\ket}[1]{|{#1}\rangle}

\newcommand{\bopk}[3]{\langle{#1}|{#2}|{#3}\rangle}

\newcommand{\realsum}{\displaystyle\sum}

\newcommand{\figref}[1]{Fig.\ \ref{#1}}

\begin{document}

\title{Engineering Many-Body Dynamics with Quantum Light Potentials and Measurements}

\author{T. J. Elliott}
\email{thomas.elliott@physics.ox.ac.uk}
\affiliation{Department of Physics, Clarendon Laboratory, University of Oxford, Parks Road, Oxford OX1 3PU, United Kingdom}
\author{I. B. Mekhov}
\affiliation{Department of Physics, Clarendon Laboratory, University of Oxford, Parks Road, Oxford OX1 3PU, United Kingdom}

\date{\today}

\begin{abstract}
Interactions between many-body atomic systems in optical lattices and light in cavities induce long-range and correlated atomic dynamics beyond the standard Bose-Hubbard model, due to the global nature of the light modes. We characterise these processes, and show that uniting such phenomena with dynamical constraints enforced by the backaction resultant from strong light measurement leads to a synergy that enables the atomic dynamics to be tailored, based on the particular optical geometry, exploiting the additional structure imparted by the quantum light field. This leads to a range of novel, tunable effects such as long-range density-density interactions, perfectly-correlated atomic tunnelling, superexchange, and effective pair processes. We further show that this provides a framework for enhancing quantum simulations to include such long-range and correlated processes, including reservoir models and dynamical global gauge fields.
\end{abstract}
\maketitle 

\section{Introduction}
The study of quantum gases trapped and controlled by optical potentials has expanded rapidly in recent years \cite{bloch2008,lewenstein2012}, as they provide a clean and versatile way to realise and observe many-body quantum dynamics, enabling quantum simulation of models from condensed matter and particle physics, and beyond. In parallel, studies of quantum light, such as cavity quantum-electrodynamics \cite{haroche2006}, have yielded fascinating results, including controlled state preparation and quantum non-demolition measurement.  Uniting these fields \cite{mekhov2012, ritsch2013} broadens both, and goes beyond the cases when either the light or matter are treated classically. Experimental \cite{baumann2010, wolke2012, schmidt2014, landig2015, klinder2015, landig2016} and theoretical works in this regime have revealed many interesting phenomena, such as the preparation of atomic states and dynamics \cite{mekhov2009b, chen2009a, mekhov2011, pedersen2014, lee2014, kollath2015, elliott2015, mazzucchi2016}, non-destructive measurement \cite{javanainen2003, mekhov2007, kozlowski2015, elliott2015b, caballero2015a}, many-body light-matter entanglement \cite{elliott2015b}, self-organisation, and other new quantum phases \cite{larson2008, maschler2008, chen2009b, gopalakrishnan2009, fernandez2010, strack2011, piazza2013, habibian2013, padhi2014, bakhtiari2015, caballero2015, ostermann2015}. 

In the aforementioned works, these effects occur due to either the collective behavior arising from the cavity-mediated interactions, or the suppression of atomic dynamics by light measurement backaction. We go beyond this, and for the first time study the union of these mechanisms. In doing so, we show that their interplay enables a selective engineering of the cavity-mediated processes, which may then be used to orchestrate dynamics for quantum simulation purposes. 

\begin{figure}
\centering
\includegraphics[width=\linewidth]{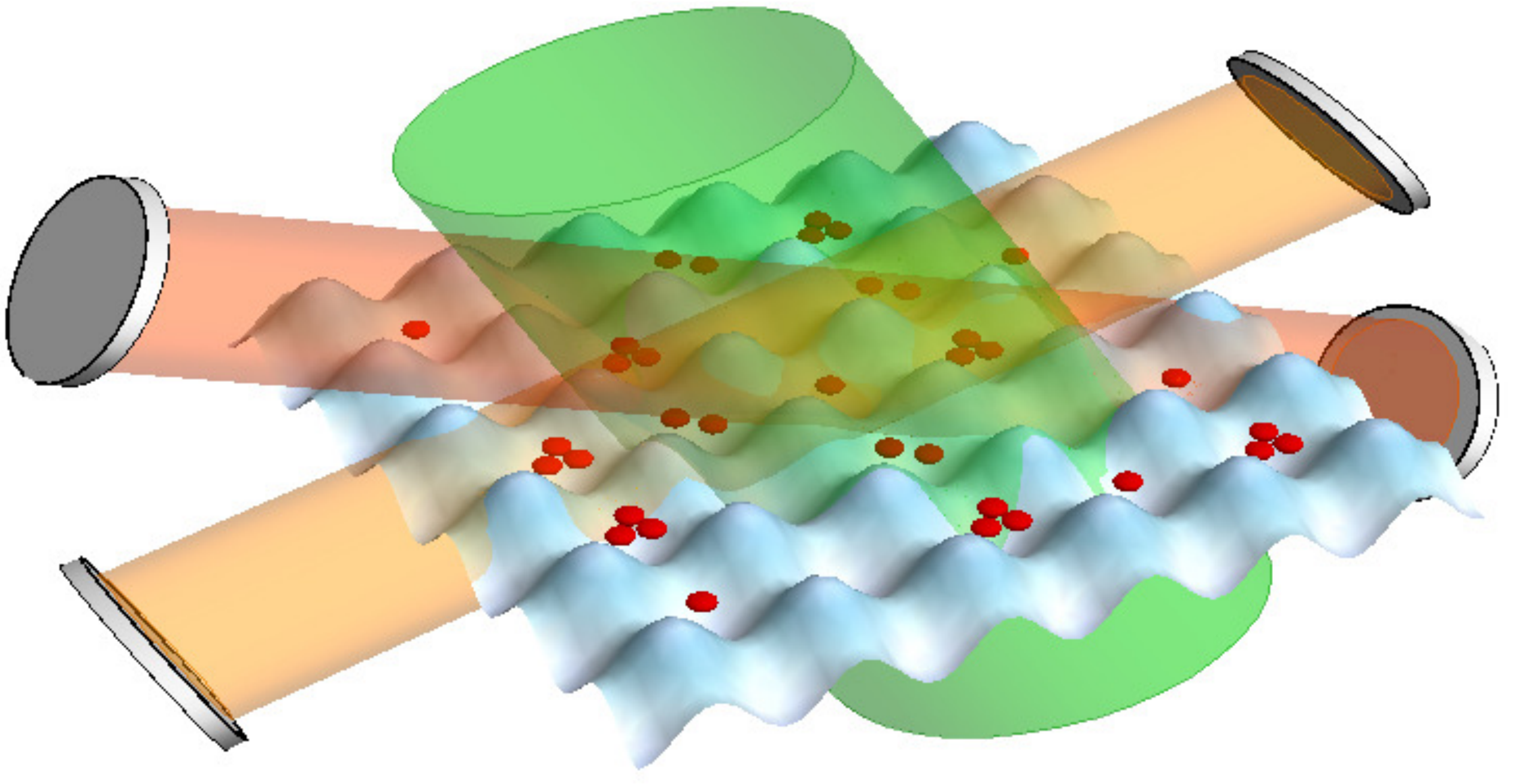}
\caption{Ultracold atoms trapped in an optical lattice scatter incident light into optical cavities. The cavity-mediated light-matter interactions induce correlated atomic dynamics that are tunable through the optical geometry. Detection of the leaked photons enables a selective suppression of these processes, through  the measurement backaction effect. This allows the atomic dynamics to be engineered for quantum simulation purposes.}
\label{figsetup}
\end{figure}

Specifically, we consider a system of ultracold (bosonic) atoms trapped in an optical lattice, probed by light. The introduction of optical cavities [\figref{figsetup}] enhances the light scattering from the atoms, and these cavities, once populated, can drive atomic dynamics. The light fields inside the cavities are dynamical quantum fields, and hence form a quantum potential for the atoms. By engineering the light modefunctions, atomic dynamics beyond the standard Bose-Hubbard model can then be realised. 

In this article, we begin by investigating the form of these dynamics, and provide a characterisation of the constituent terms. We show that this classification reveals processes that include perfectly correlated tunnelling, effective pair creation and annihilation, long-range tunnelling, superexchange, and independently tunable long- and short-range density-density interactions. We illustrate how these different effects may be controlled and tuned by the optical geometry.

Following this, we introduce the backaction effect that arises from measurement of the light leaking from the cavity to these extended dynamics. We demonstrate how this enables further control of the processes, by imparting additional structure to the lattice, allowing for the highlighting of desired effects (by suppression of others) through quantum Zeno dynamics \cite{facchi2008}. We describe how this united formalism provides a framework to enhance quantum simulations, through the incorporation of these correlated and long-range processes that are not accessible in other systems with finite-range interactions, through the introduction of building block components, which include reservoir models and dynamical global gauge fields.

\section{Cavity-Mediated Dynamics}

\subsection{The Model}
\label{secmodel}

In this article we study a extended form of the Bose-Hubbard model \cite{mekhov2012} in which interactions with an additional set of modes possessing long-range spatial extent over the lattice are included. Physically, this can correspond to the scenario where the lattice is a (classical) optical trap containing bosonic atoms, embedded within an optical cavity, with light from an external laser scattered by the atoms into the cavity modes. A key feature of this model is that both the light and the atoms are dynamical quantum fields. It forms the cornerstone of many of the aforementioned works in the field of fully-quantum many-body light-matter interactions, and in its general form describes a wide range of possible optical geometries. We outline the main steps in the derivation of the effective atomic Hamiltonian; more detailed treatments may be found in, e.g.~\cite{maschler2008, mekhov2012, ritsch2013, caballero2015, caballero2015a}.

The full Hamiltonian can be written $\mathcal{H}=H_L+H_M+H_{LM}$, where (in natural units) 
\begin{subequations}
\begin{equation}
H_L=\sum_m \omega_m a^\dagger_ma_m
\end{equation}
and
\begin{equation}
H_M=\sum_{ij}J_{ij}^Tb^\dagger_ib_j+\frac{U}{2}\sum_{i}b^\dagger_ib^\dagger_ib_ib_i
\end{equation}
\end{subequations}
are the bare Hamiltonians describing the light and matter respectively (the matter Hamiltonian being the standard Bose-Hubbard Hamiltonian for atoms in a lattice when $J^T_{ij}$ is restricted to nearest-neighbour terms only). Here, $a^\dagger_m$ creates photons in light mode $m$ with frequency $\omega_m$ and modefunction $u_m(\bm{r})$, while $b_i^\dagger$ creates bosons at lattice site $i$ with Wannier function $w(\bm{r}-\bm{r}_i)$. On-site interactions between atoms are parameterised by $U$, and $J_{ij}^T$ are the (classical) tunnelling rates between sites due to the classical potential.

The final term $H_{LM}$ describes the fully quantum light-matter interactions between the light and atomic modes. It follows from a many-atom generalisation of the Jaynes-Cummings Hamiltonian, with the excited atomic state adiabatically eliminated \cite{mekhov2012}. To obtain this, consider first the single-particle Hamiltonian for an atom interacting with many light modes: the interaction part of the Hamiltonian can be written 
\begin{equation}
H_{\mathrm{LM}}^{\mathrm{(SP)}}=\sum_m g_ma_mu_m(\bm{r})\sigma^++h.c.,
\end{equation}
where $\sigma^+$ raises the atom to its excited state, and $g_m$ is the light-matter coupling constant for mode $m$. To perform the adiabatic elimination, we assume that the detuning is sufficiently large that the excited state population is negligible, and from the Heisenberg equation $\dot{\sigma}^-=i[H^{\mathrm{(SP)}},\sigma^-]$ set the time dependence of this operator to vanish in a frame rotating at a reference frequency $\omega_p$ (e.g.~that of an external pump laser) leading to $\sigma^-=(1/\Delta_a)\sum_mg_mu_m(\bm{r})a_m$, where $\Delta_a=\omega_p-\omega_a$ is the atomic transition frequency $\omega_a$ detuning from $\omega_p$. Inserting this into the single-particle Hamiltonian, we have 
\begin{equation}
H_{\mathrm{LM}}^{\mathrm{(SP)}}=\sum_{mn}\frac{g_mg_n}{\Delta_a}u_m^*(\bm{r})u_n(\bm{r})a^\dagger_ma_n.
\end{equation}
 Finally, to obtain the many-body form of the Hamiltonian, we express the single-particle state in terms of the localised atomic basis states $\psi(\bm{r})=\sum_iw(\bm{r}-\bm{r}_i)b_i$, resulting in \cite{mekhov2012}

\begin{equation}
\label{eqHlm}
H_{LM}=\sum_{mn}\frac{g_mg_n}{\Delta_a}a^\dagger_ma_n\sum_{ij}J_{ij}^{mn}b_i^\dagger b_j.
\end{equation}

The interactions are parameterised by 
\begin{equation}
J_{ij}^{mn}=\int d\bm{r} w(\bm{r}-\bm{r}_i)u^*_m(\bm{r})u_n(\bm{r})w(\bm{r}-\bm{r}_j),
\end{equation}
describing the overlap between the atomic Wannier functions and light modefunctions. These coefficients thus encompass the dependence of the dynamics on the particular optical setup used in the system. We note also that due to the complex nature of light modefunctions, these coefficients may too be complex. Taking the Wannier functions to be real-valued, these coefficients satisfy the properties $J_{ij}^{mn}=J_{ji}^{mn}={J_{ij}^{nm}}^*$.

Let us now consider one of the light modes, which we label $0$, to be a pump mode sourced from an external laser. We describe this mode by a coherent state of amplitude $\alpha_0$, with an occupation much larger than the cavity modes, such that we can replace $a_0\to\alpha_0$. Assuming that the light scattering occurs on timescales much faster than the atomic dynamics, we can obtain the time dependence of the cavity modes (in the frame rotating at the pump frequency) from the Heisenberg equation: 
\begin{align}
\dot{a}_{m\neq0}&=i[\mathcal{H},a_{m\neq0}]\nonumber \\ &=-i\Delta_ma_m-i\sum_{n}\Omega_{mn}\mathcal{J}_{mn}-\kappa_ma_m.
\end{align}
In this expression, the cavity detuning $\Delta_m=\omega_p-\omega_m$, and we have phenomenologically introduced a cavity decay for photon loss with rate $\kappa_m$, and defined for shorthand $\Omega_{mn}=g_mg_n/\Delta_a$ and 
\begin{equation}
\mathcal{J}_{mn}=\sum_{ij}J_{ij}^{mn}b_i^\dagger b_j.
\end{equation}
These light-matter coupling operators $\mathcal{J}_{mn}$ inherit certain properties from the coupling coefficients $J_{ij}^{lm}$: $\mathcal{J}_{mn}=\mathcal{J}_{nm}^\dagger=\mathcal{J}_{nm}^*$.

We now further assume also that there is only a small dispersive frequency shift of the cavity modes $(\Omega_{mm}\mathcal{J}_{mm}\ll\Delta m$ for $m\neq0$), and use that $\Omega_{m0}\alpha_0\mathcal{J}_{m0}\gg\sum_{n\neq\{0,m\}}\Omega_{mn}\mathcal{J}_{mn}$ due to the large pump amplitude compared to the cavity occupations. Thus, the steady states of the cavity modes are hence
\begin{equation}
\label{eqsteady}
a_{m\neq0}=\frac{\Omega_{m0}\alpha_0\mathcal{J}_{m0}}{i\kappa_m+\Delta_m}\equiv C_m\mathcal{J}_{m0}.
\end{equation}

When these steady states are reached on timescales much faster than the atomic dynamics (i.e.~$\kappa_m\gg J^T_{ij}$), we can perform a further adiabatic elimination, to remove the cavity modes from the Hamiltonian. Replacing the cavity mode operators with their steady state values in $\mathcal{H}$ results in the effective atomic Hamiltonian \cite{maschler2008,caballero2015}
\begin{align}
\label{eqheff}
\mathcal{H}=&H_M+\Omega_{00}|\alpha_0|^2\mathcal{J}_{00} \nonumber \\ 
&+\realsum_{m\neq0}\frac{\Delta_m|C_m|^2}{2}(\mathcal{J}_{m0}^\dagger\mathcal{J}_{m0}+\mathcal{J}_{m0}\mathcal{J}_{m0}^\dagger). 
\end{align}
Note that the terms in Eq.~\eqref{eqHlm} containing products of two cavity modes are neglected, as they are much smaller than the pump-pump and pump-cavity product terms. Note also that the symmetric splitting of the products of $\mathcal{J}_{m0}$ and its conjugate into two parts of opposite order are necessary to preserve the form of the Heisenberg equations for the atomic modes $b_j$ before and after the elimination, as the ordering freedom of $a_m$ and $b_j$ is lost after the steady state replacement (see Appendix) \cite{maschler2008}. This regime in which the cavity steady state is reached much faster than the timescales of atomic dynamics is readily accessible in experiments, and indeed has already been demonstrated \cite{landig2016}; typical tunnelling rates $J^T_{ij}$ between nearest-neighbour sites are $\mathcal{O}(10^3$Hz) \cite{jaksch1998}, while cavities with decay rates $\kappa_m$ $\mathcal{O}(10^6$Hz) are found in experimental use \cite{landig2016}.

Beyond the standard Bose-Hubbard Hamiltonian, the first additional term in this effective Hamiltonian is the pump-pump term, due to the interaction between the atoms and the (classical) light from the pump laser. Such terms can be derived straightforwardly from semi-classical treatments of light-matter interactions, and is essentially of the form of a Raman transition, giving rise to light-induced tunnelling and effective chemical potentials. The inclusion of additional pumps will manifest similar such terms, including terms mixing different pump modes, which have previously been used to introduce complex phases to atomic tunnelling (e.g.~\cite{jaksch2003}). The second set of additional terms, the cavity-pump terms, arise due to interaction of the atoms with the quantised cavity light fields, and do not appear in semi-classical treatments, as they are inherently defined by the backaction of the atomic state on the cavity population. These terms, due to the long-range spatial extent of the cavity modes, induce effective long-range correlations and interactions between lattice sites. We shall primarily focus on the use of these two-body dynamical processes.

The choice of light modefunctions leads to different dynamics, and are tunable by a variety of methods, including changing the wavelength and angle of the lasers, and the angle and size of the cavities. We highlight that multimode, or even multiple cavities allow flexibility beyond a single cavity mode. We now proceed to characterise the induced processes, focussing on the two dominant contributions, from the on-site, and neighbouring inter-site terms. In contrast to previous works investigating these additional terms, which treat them as a whole and consider them generically as four-point correlations \cite{caballero2015}, we introduce a characterisation that becomes meaningful and important when we later introduce a method to distinguish between and selectively tailor such processes, through the measurement backaction.

\subsection{On-site Terms}
The on-site terms will typically dominate in the light-matter interaction operators $\mathcal{J}_{mn}$ \cite{mekhov2012}, and thus these operators can often be approximated with replacement by their on-site counterparts 
\begin{equation}
D_{mn}=\sum_i J_{ii}^{mn}b^\dagger_i b_i.
\end{equation}
Due to the perfect overlap of the Wannier functions (as they are identical), the corresponding light-matter interaction coefficients $J_{ii}^{m0}$ have close to unit magnitude when the light modefunctions are at peak intensity at the centre of lattice sites \cite{mekhov2012, caballero2015}. These coefficients may be imprinted with complex phases through the phase difference of the incoming and outgoing light modefunctions. This allows for the generation of a matter mode structure, in which the lattice is partitioned into sets of sites scattering light with the same phase, and the modes are defined by atoms occupying these sets of sites in which they scatter light with the same phase \cite{elliott2015}. For example, when considering the main diffraction maximum, we have that $J_{ii}^{m0}=(-1)^i$, and hence all odd sites scatter light with the same phase (thus forming one matter mode), while all even sites scatter light with the same, opposing phase (hence forming the second matter mode).

Focussing first on the special case of illumination in the diffraction maximum, in which the light-matter interaction coefficients are all identically $J_{ii}^{m0}=1$ (and similarly, $J_{ii}^{00}=1$) the interaction terms become $D_{m0}=N_m$, where $N_m$ is the number of atoms in total that occupy any of the sites illuminated by both the pump, and cavity mode $m$. Analogously, we have that $D_{00}=N_0$. Thus, the light-induced dynamics in the effective atomic Hamiltonian Eq.~\eqref{eqheff} is given by
\begin{equation}
\label{eqdensdens}
\Omega_{00}|\alpha_0|^2N_0+\sum_{m\neq0}\Delta_m|C_m|^2N_m^2,
\end{equation}
where the first term forms an effective chemical potential, and the second set of terms mediate density-density interactions between sites illuminated by the pump and their respective cavity mode. These latter interactions occur irrespective of the spatial separation of the sites, and thus exemplify the long-range nature of the cavity-induced processes.

\begin{figure}
\centering
\includegraphics[width=\linewidth]{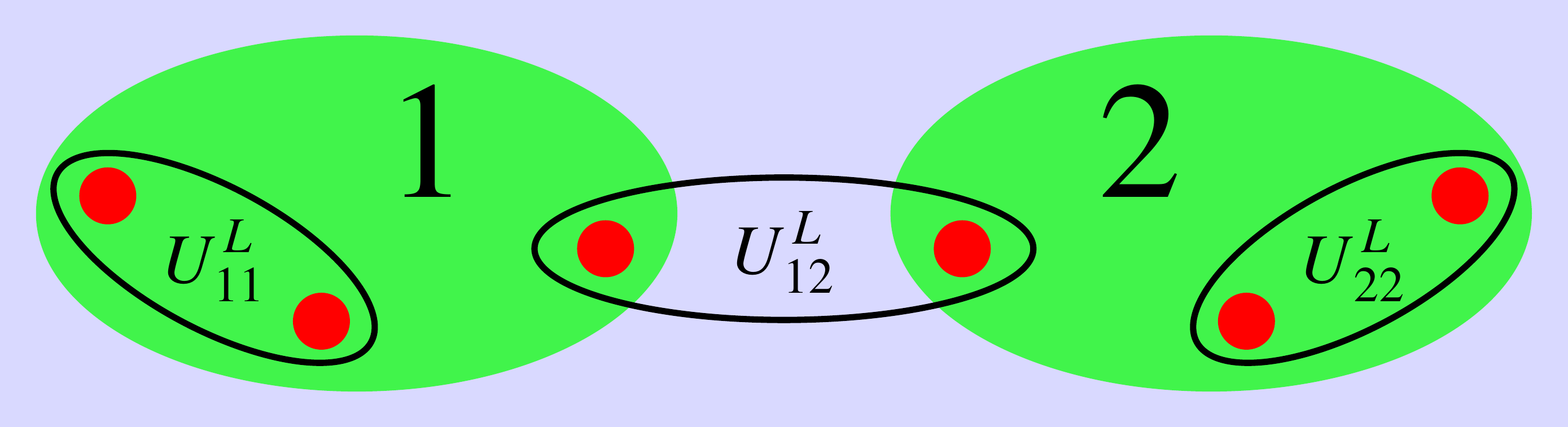}
\caption{Schematic displaying the regions and associated interaction strengths of cavity-mediated density-density interactions. The flexibility in the geometry of the cavity modes allows for the strength of the long- and short-range processes to be tuned independently.}
\label{figregions}
\end{figure}

When considering the inclusion of multiple cavity modes, the tunability of these interactions will exceed what is possible with a single cavity. One such possibility this provides is that the long- and short-range interaction strengths may be tuned independently, as can be seen by noting that the sign of the cavity detuning determines whether the long-range density-density interactions are repulsive or attractive, and thus different cavities can have different contributions to the overall dynamics.

As can be seen from Eq.~\eqref{eqdensdens}, a single pump and cavity mode will produce density-density interactions with a strength $U^L=\Delta_m|C_m|^2$ between atoms on all sites illuminated by pump and mode $m$. An illustration of how short- and long-range interactions between two regions may be varied can be seen by considering three cavity modes, which we label $X$, $Y$, and $Z$, which illuminate regions 1, 2, and both respectively (see \figref{figregions}) at the diffraction maximum, with both regions illuminated by a common pump. Denoting the light-mediated density-density interaction strength between an atom in region $A$ and an atom in region $B$ as $U^L_{AB}$, we hence have
\begin{subequations}
\begin{equation}
U_{11}^L=\Delta_X|C_X|^2+\Delta_Z|C_Z|^2,
\end{equation}
\begin{equation}
U_{22}^L=\Delta_Y|C_Y|^2+\Delta_Z|C_Z|^2,
\end{equation}
and
\begin{equation}
U_{12}^L=\Delta_Z|C_Z|^2.
\end{equation}
\end{subequations}

Thus, the three interaction strengths can all be tuned independently of each other, through the respective $C_m$ and $\Delta_m$ of each cavity mode. We note also that in general, as the regions are defined by the matter mode structure, which is in turn defined by the light-matter interaction coefficients $J_{ii}^{m0}$, the regions considered here need not be spatially contiguous, and indeed, the modes can have a very non-trivial spatial overlap with each other \cite{elliott2015}. In this sense, one can more generally consider the interaction strengths as being classed as inter- and intra-mode, rather than short- and long-range.

For more general, arbitrary illumination patterns, all of the sites in the illuminated region will still be part of the resultant density-density interactions. The interaction strength between two particular sites (or modes) is determined by their $J_{ii}^{mn}$, and for a cavity mode $c$, the associated interaction strength between modes $A$ and $B$ is
\begin{equation}
U_{AB}^L=\Delta_c|C_c|^2\cos(\phi_{AB}),
\end{equation}
where $\phi_{AB}$ is the difference between the phases of the associated $J_{ii}^{m0}$ of the two modes. For multiple cavity modes, one sums up the contributions from each cavity individually, while with multiple pumps one must sum over the $C_c$ generated by each pump, and also consider the additional pump-pump terms in the chemical potential, which may now differ from unit strength.

The effective atomic interactions resulting from such cavity-induced processes have already been demonstrated experimentally for the special case of illumination at the main diffraction minimum \cite{landig2016}. In these experiments, the resulting interactions were of the form $\Delta_c|C_c|^2(N_c^{(\mathrm{even})}-N_c^{(\mathrm{odd})})$, with the detuning chosen such that this term favours a population imbalance between even and odd states, leading to cavity-induced atomic self-organisation for large enough interaction strengths. The experiment demonstrated that this interaction strength can be made comparable to, and even stronger than the tunnelling from the standard Bose-Hubbard model by one or two orders of magnitude (i.e.~up to $\mathcal{O}(10^4-10^5$Hz))\cite{landig2016}. As the other more general schemes based on using the on-site terms in the light-matter interaction operator typically involve only changing the phase of the $J_{ii}^{mn}$ through adjustment of the optical geometry, the size of such interaction strengths can be expected to remain of a similar size.

\subsection{Inter-site Terms} 

In contrast to the on-site case discussed above, when the overlap of the light modes are arranged to be concentrated between lattice sites the nearest-neighbour terms in $\mathcal{J}_{mn}$ can be made more significant than the on-site terms \cite{kozlowski2015}. Specifically, when the modefunction of cavity $c$ and the pump modefunction are given by standing waves with wavenumbers $k_c=k_0=\pi/d$ ($d$ being the lattice spacing), at opposing angles to the lattice $\theta_0=-\theta_c$, then the on-site pump-cavity light-matter coupling coefficients vanish ($J^{c0}_{ii}=0$), while the nearest-neighbour coefficients $J^{c0}_{\mathrm{nn}}$ (the subscript nn denoting nearest-neighbour sites) take a constant value at all site pairs with bonds in a given direction (see \cite{kozlowski2015} for further details). Intuitively, this can be seen to occur because the Wannier functions of a site are symmetric, while the product of light modefunctions $u_c^*(\bm{r})u_0(\bm{r})$ are antisymmetric, and periodic across two lattice sites, hence leading to a cancellation of their overlap with each on-site product of Wannier functions, but not the inter-site products. Note that because $|u_0(\bm{r})|^2$ is a positive-definite function, such a cancellation does not occur for the $J_{ii}^{00}$; however, for such an illumination pattern, this coefficient is equal for all illuminated lattice sites, forming a constant effective potential within this region, and hence when the pump illuminates the entire lattice this may generally be neglected in the effective Hamiltonian Eq.~\eqref{eqheff}. The pump-pump inter-site coefficients $J_{\mathrm{nn}}^{00}$ will then also take a constant value.

Using this, in this regime we replace the light-matter coupling operators by these inter-site terms alone:
\begin{equation}
B_{mn}=\sum_{<ij>} J_{ij}^{mn}b^\dagger_ib_j,
\end{equation} 
where $<ij>$ indicate neighbouring site pairs. These terms describe light-induced atomic tunnelling events. While in general the $J_{<ij>}^{mn}$ are different for each pair of sites, leading to spatially-dependent tunnelling rates, which may be tuned as with the density-density interactions, we shall here focus on the aforementioned case where they are homogeneous. This allows for the one-body tunnelling terms (as would be expected from semiclassical treatments) to be suppressed or enhanced by the pump light, or potentially even eliminated. This latter case would leave only the two-body terms, the rates of which may be tuned semi-independently of the one-body terms.

\begin{figure}
\centering
\includegraphics[width=\linewidth]{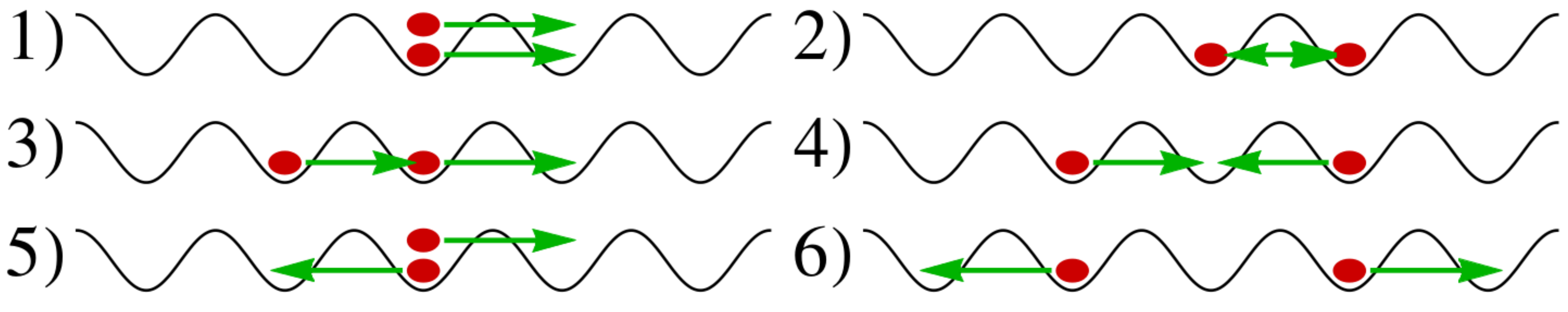}
\caption{The cavity fields can also induce tunable correlated tunnelling processes, which may be classified according to the relationship between the two tunnelling atoms: (1) pair tunnelling; (2) pair exchange; (3) effective next-nearest-neighbour tunnelling; (4 and 5) effective pair processes; and (6) general long-range correlated tunnelling.}
\label{figtunnelling}
\end{figure}

These two-body processes are of the form 
\begin{equation}
\label{eqtwobodyproc}
\Delta_m\frac{|C_m|^2}{2}\sum_{\substack{<ij>\\<kl>}} (J_{ij}^{m0}J_{kl}^{0m}b^\dagger_ib_jb^\dagger_kb_l+h.c.),
\end{equation}
where $\{i,j\}$ and $\{k,l\}$ must be pairs of neighbouring sites, though the two pairs may be distributed anywhere within the illuminated regions. Each pair corresponds to a tunnelling process, and hence the complete two-body processes correspond to correlated tunnelling events. These can be classified according to the relationship between the two site pairs [\figref{figtunnelling}], giving rise to (1) pair tunnelling, (2) pair exchange, (3) effective next-nearest-neighbour tunnelling, (4 and 5) effective pair processes, and (6) general long-range correlated tunnelling. Due to the smaller inter-site overlap of Wannier functions compared to on-site overlaps, the inter-site light-matter coupling coefficients will necessarily be smaller than the  on-site coefficients at their respective maxima. Their relative magnitude has previously been studied in \cite{caballero2015}, where they are shown to typically be separated by approximately an order of magnitude. Thus, the complete two-body correlated tunnelling terms can have a strength approximately two orders of magnitude less than that which may be achieved for the light-induced density-density interactions, and hence can occur at rates comparable to the classical tunnelling $J_{\mathrm{nn}}^{T}$ (i.e.~$\mathcal{O}(10^3$Hz)). 

\subsection{Simulation of Superexchange Interactions}

Before we introduce the measurement backaction as a method of control for these processes, we shall first suggest an alternative approach for exerting additional tunability beyond the optical geometry. One such way is to impart an additional structure to the atomic system by shaping the underlying lattice, so that it is no longer homogeneous at every site. We provide as an example of this a proposal for how superexchange interactions in spin models may be simulated using the setup.

Two-body atomic tunnelling processes have previously been used to implement such simulations \cite{trotzky2008}. However, in these earlier proposals, the superexchange occurs as a second-order perturbative process. In contrast, here we suggest using the cavity-induced pair exchange processes for the same purpose. Our proposal follows the original by dividing the lattice into pairs of double wells with a superlattice potential, with each site pair containing two atoms of different (pseudo)spin species, and strong interparticle interactions enforcing a one-particle-per-site constraint (we also assume the use of the same method for the initial state preparation). Introducing the correlated tunnelling as above for the inter-site case (through use of a single cavity mode $c$), the superlattice potential and limit on site occupation effects constraints on the allowed dynamical processes, permitting only pair exchange processes between atoms in each site pair.

\begin{figure}
\centering
\includegraphics[width=\linewidth]{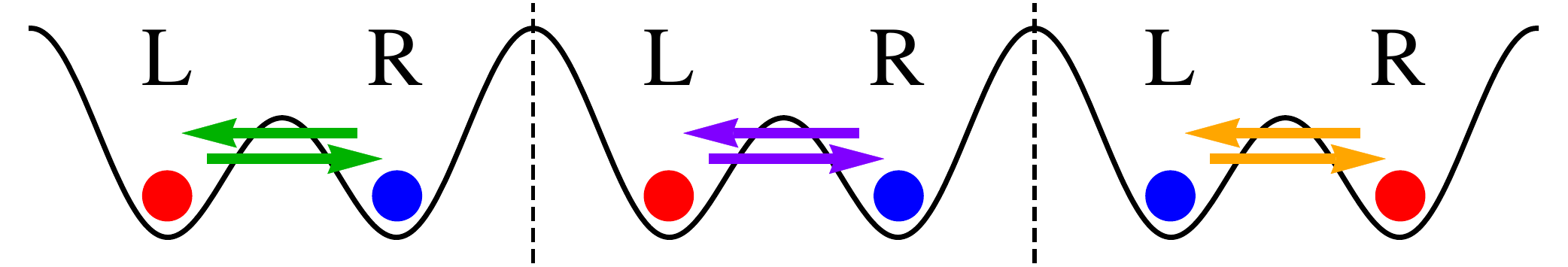}
\caption{The correlated tunnelling processes may be used to simulate superexchange interactions. Here, a superlattice is used to divide the system into a series of double well potentials, each containing an atom of each of two spin species (denoted by their colour). The superlattice potential and strong repulsive interactions allow only pair exchange processes to take place, between atoms in the same double well, mimicking superexchange dynamics. These correlated tunnellings are indicated by arrows of the same colour.} 
\label{figsuperexchange}
\end{figure}

To see this, consider each of the possible tunnelling events. The superlattice potential only permits tunnelling of the atoms into the corresponding other site in the double well pair. The one-particle-per-site constraint suppresses any process in which an atom tunnels to an already occupied site (as is the case in each of the double wells). However, when we consider the correlated processes, such tunnelling events can take place when coupled with another tunnelling that preserves the unit occupation of each site. There are two such processes: those in which the atom of the other species in the site pair tunnels in the opposite direction, and those in which the same atom tunnels in the reverse direction (see \figref{figsuperexchange}). The latter processes take the form $b^\dagger_{Lx}b_{Rx}b_{Rx}^\dagger b_{Lx}$, where $\{L,R\}$ denotes the corresponding site in the well, and $x$ the spin species. Due to the unit occupation of each site pair by each spin species, these terms have a constant value, and so may be neglected. 

Thus, each double well (up to constant terms) behaves according to the Hamiltonian 
\begin{equation}
H_{\mathrm{ex}}=2\Delta_c|C_cJ_{\mathrm{nn}}^{c0}|^2(b_{L\uparrow}^\dagger b_{R\uparrow} b^\dagger_{R\downarrow}b_{L\downarrow}+h.c.).
\end{equation}
This can equivalently be expressed in terms of spin operators ($2S_j^Z=b^\dagger_{j\uparrow}b_{j\uparrow}-b^\dagger_{j\downarrow}b_{j\downarrow}$), to give
\begin{equation}
H_{\mathrm{ex}}=J_{\mathrm{ex}}(S_L^+S_R^-+S_L^-S_R^+),
\end{equation}
where 
\begin{equation}
J_{\mathrm{ex}}=2\Delta_c|C_cJ_{\mathrm{nn}}^{c0}|^2.
\end{equation}
Critically, the exchange term here does not suffer the $1/U$ dependence of the second-order process in the original proposal, and hence the interparticle interactions necessary for enforcing the single-particle-per-site constraint can here be increased without suppressing the exchange interaction. Indeed, as noted above, these correlated tunnelling processes can occur at rates comparable to the standard tunnelling $J^T_{\mathrm{nn}}$ from the classical optical lattice potential.

\section{Inclusion of the Light Measurement Backaction}

A drawback of the above method of imparting additional structure by altering the lattice geometry is that it bears an adverse consequence whereby the suppression of a particular tunnelling event will also necessitate such a suppression when it would otherwise have occurred as part of a correlated tunnelling event. For example, if the process $b^\dagger_Xb_Y$ is suppressed by the lattice geometry causing it to occur with low amplitude, then any process of the form $b^\dagger_Xb_Yb_i^\dagger b_j$ also has a correspondingly low amplitude. In contrast to this, the matter mode structure discussed above offers an alternative avenue for imprinting further structure onto the atoms, without having to suffer such a penalty.

In previous works \cite{elliott2015, mazzucchi2016}, we have discussed how the light measurement backaction from the photons leaked by a cavity can be used to selectively suppress atomic dynamics in the standard Bose-Hubbard model, through constraints on the dynamics imposed by the rate at which photons are detected. We now go beyond this, and incorporate the cavity light-induced dynamics into such methods, thus uniting the measurement backaction effect with the cavity backaction for the first time. We will then evince the potential of this union for enhancing quantum simulations with atomic systems. Crucially, the measurement backaction allows for a process to be forbidden as a single event, but permitted when correlated with another particular event (or events) commensurate with the measurement outcome; in this case the suppression is achieved through constraints on the matter mode occupations.

This measurement backaction is realised through the introduction of an additional cavity (and possibly pump) mode, where measurement is made of the scattered light that it leaks. The cavity leaks photons at a rate that is proportional to its occupation, which is determined by the particular atomic state through Eq.~\eqref{eqsteady}. We will here assume that this measurement cavity and associated pump is arranged such that the on-site terms dominate the light-matter interactions, and hence for a given atomic Fock state configuration $\bm{n}$, the cavity mode $\Pi$ has a well-defined amplitude 
\begin{equation}
a_\Pi=C_\Pi D_{\Pi0}=C_\Pi\sum_j J_{jj}^{\Pi0} n_j.
\end{equation}
For such a configuration, the (average) rate of photon leakage from the cavity is constant. More generally, the atomic configuration is a superposition of such Fock states. Consider a state where the Fock states $\bm{n}$ occur with initial amplitudes $c_{\bm{n}}^0$. Applying the quantum jump measurement formalism, these amplitudes then evolve according to \cite{mekhov2012}
\begin{equation}
c_{\bm{n}}(k,t)=\frac{1}{\mathcal{N}}\alpha_{\Pi\bm{n}}^ke^{-|\alpha_{\Pi\bm{n}}|^2\kappa t}c_{\bm{n}}^0
\end{equation}
where $\alpha_{\Pi\bm{n}}=C_\Pi\bopk{\bm{n}}{D_{\Pi0}}{\bm{n}}$ is the cavity field amplitude for the given Fock state $\bm{n}$, and $\mathcal{N}$ is a normalisation factor. In this expression, the first factor represents the effect of the quantum jumps occurring at each of the $k$ photodetection events of the leaked photons, while the second factor gives the non-Hermitian evolution occurring between such jumps during the elapsed time $t$. This evolution then enacts a natural selection of sorts, reducing the relative probabilities of states not consistent with the observed leakage rate. For intense, persistent measurement of this form, the distribution of amplitudes is compressed, and consequently the light field state converges towards a particular coherent state $\alpha_{\Pi z}$. When this convergence to a single state happens on timescales much shorter than the atomic dynamics, the light field is ultimately pinned to this state (through the quantum Zeno effect \cite{misra1977}). The requisite condition on the timescales can be expressed as $|C_\Pi|^2\kappa_\Pi\gg J^T_{\mathrm{nn}}$ \cite{mazzucchi2016}. Note that while this regime is not reached in the aforementioned example experiment \cite{landig2016} (because their intent was to study the cavity backaction alone), it would be feasible by, e.g.~a modest reduction of the cavity detuning, or an increase in the pump power, both of which have been performed in previous incarnations of the same setup \cite{baumann2010}.

 The matter is then confined to evolve within only a subspace of its full Hilbert space, which is determined by the particular $\alpha_{\Pi z}$; specifically, it must remain within the subspace of states $\{\bm{n}_z\}$ for which $\alpha_{\Pi\bm{n}_z}=\alpha_{\Pi z}$. This set of states is defined by the matter mode structure, and hence depends on the modefunctions of the measurement cavity and pump \cite{elliott2015}. The ensuing atomic dynamics must take place within this measurement subspace, thus undergoing quantum Zeno dynamics \cite{facchi2008}. In this regime, the dynamics is described by the appropriate Zeno Hamiltonian, defined as
\begin{equation}
H_Z=\mathcal{P}H\mathcal{P},
\end{equation}
where $\mathcal{P}$ is the projector that describes the subspace of states $\{\bm{n}_z\}$ consistent with the measured light state \cite{facchi2008}. This result follows generally from considerations of a system subject to very frequent measurement: in the limit that the number of measurements $N\to\infty$ in a fixed time $t$, the evolution $(\mathcal{P}\exp(-iHt/N))^N$ can be expanded approximately as \cite{facchi2008}
\begin{align}
\lim_{N\to\infty}(\mathcal{P}e^{-iH\frac{t}{N}})^N&\approx\lim_{N\to\infty}(\mathcal{P}(1-iH\frac{t}{N}))^N\nonumber\\&=e^{-iH_Zt},
\end{align}
with the second line following from the definition of the exponential function $\exp(x)\equiv\lim_{n\to\infty}(1+x/n)^n$, and the idempotence of the projector ($\mathcal{P}^2=\mathcal{P}$). We shall now drop the $Z$ subscript from the Zeno Hamiltonians for the remainder of this article.

Thus, utilising this formalism, we can use the measurement backaction to selectively eliminate particular atomic dynamics, as determined by the measurement cavity geometry and the associated projection operators. In contrast to earlier work incorporating measurement backaction, in which the two-body terms that appear due to measurement are only second-order processes \cite{mazzucchi2016}, in this scheme such terms now arise here at first order in the system evolution, as they are directly induced by the cavity backaction (rather than simply being higher-order processes which are not suppressed by the measurement), and are hence perfectly correlated. The two cases are thus fundamentally different, and here increasing pump strength increases the two-body tunnelling rates, rather than suppressing them. Note that because the dynamics of the system is now constrained to the subspace of states consistent with a single value of the measurement cavity light-matter interaction operator, the dynamics induced by the measurement cavity may be disregarded, as they are identical for all states in the subspace, and hence form a constant energy shift for all states.

As a straightforward example of such a scheme, consider the case of measurement made at the diffraction minimum ($J^{\Pi0}_{ii}=(-1)^i$; $D_{\Pi 0}=N_\Pi^{\mathrm{(even)}}-N_\Pi^{\mathrm{(odd)}}$) across the lattice. This freezes the occupation number difference between odd and even sites, and thus when this difference is given by $\Delta N=N^{(\mathrm{odd})}-N^{(\mathrm{even})}$, the allowed states are superpositions of states of the form $\ket{\bm{n}^{(\mathrm{even})},\bm{n}^{(\mathrm{odd})}}$ with $\sum_{j\in\mathrm{(even)}} n_j = n$ and $\sum_{j\in\mathrm{(odd)}} n_j=n+\Delta N$, for integer $n>0$. The dynamics is restricted to this subspace of states with associated projector 
\begin{equation}
\mathcal{P}_{\Delta N}=\sum_n\ket{n,n+\Delta N}\bra{n,n+\Delta N}
\end{equation}
by the Zeno dynamics, and since single nearest neighbour tunnelling events change the $\Delta N$ of a state, such processes are forbidden from taking place by the measurement. In the absence of a further cavity driving dynamics this would leave next-nearest neighbour tunnelling as the leading process, which typically occurs at a much slower rate than nearest-neighbour tunnelling in the standard Bose-Hubbard model (and thus is often ignored), but may now be no longer negligible. 

\begin{figure}
\centering
\includegraphics[width=\linewidth]{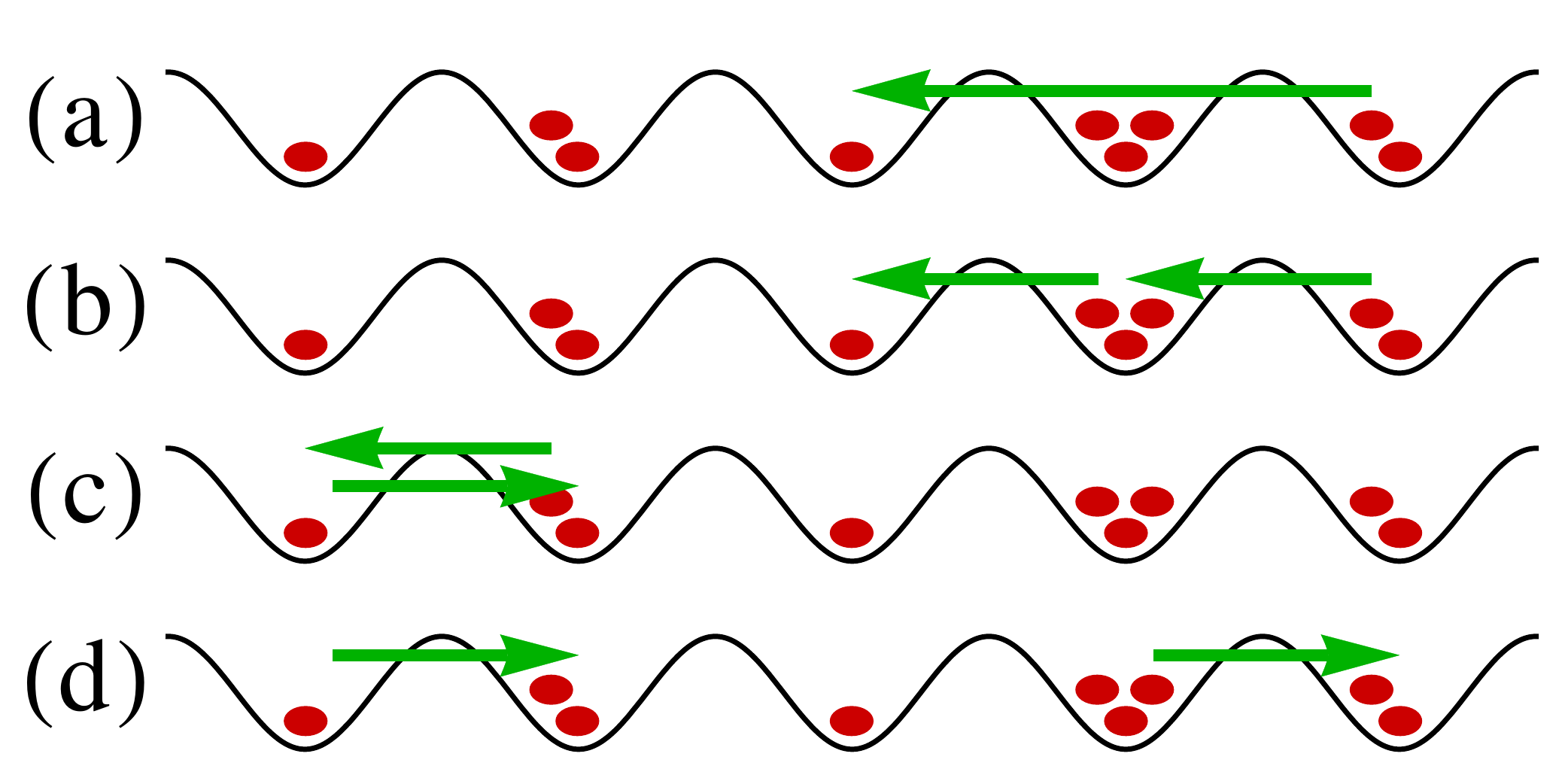}
\caption{Illustration of the allowed processes in the presence of measurement-induced dynamical constraints requiring fixed occupation number difference between even and odd sites ($\Delta N=N^{(\mathrm{odd})}-N^{(\mathrm{even})}$). These allowed processes must conserve the total occupation difference between even and odd sites, and include: (a) nearest neighbour tunnelling; (b) effective nearest-neighbour tunnelling; (c) pair exchange; and (d) long-range correlated tunnelling events that preserve total mode occupations.}
\label{figtwomode}
\end{figure}

Considering also the cavity backaction when a cavity is present to drive dynamics, this scenario may be augmented with the two-body terms introduced above (namely, those that preserve matter mode occupation number difference). The correlated processes in which two atoms tunnel into the same site (and the reverse process) are forbidden by the measurement as they violate the constraint on $\Delta N$, as are certain of the long-range correlated tunnelling events. The allowed processes of the latter form lead to effective long-range tunnelling events within each mode. All of the light-induced effective nearest-neighbour tunnelling events are permitted, as they simply move atoms between neighbouring sites in each of the two modes, as are the pair exchange events, as they do not change the site occupation numbers. \figref{figtwomode} illustrates examples of each of the allowed processes in this two mode example. In the simplest case of uniform illumination, the rate of the long-range correlated tunnelling processes are independent of the site separation, though as noted above, akin to the density-density interactions, spatially-dependent long-range tunnelling rates can be tuned with multiple cavity modes.

The flexibility of the light-modes allows a range of configurations of the two-body terms to be engineered through measurement backaction; for example, when light measurement fixes occupation numbers for multiple matter modes \cite{elliott2015}, the only allowed correlated tunnelling events are pair exchange, and the long-range which preserve the total mode occupations. 

\section{Framework for Quantum Simulations}

The architecture based on the union of measurement backaction and cavity backaction that we have detailed above naturally lends itself to quantum simulations. In particular, it offers opportunities to mimic correlated processes and long-range interactions of the atoms that would be difficult (or even not possible) to realise in systems with finite-range interactions. We now proceed by outlining a framework to this end, by describing `building block' components for implementing reservoir and dynamical global gauge field models, which can be used to enhance methods of optical lattice quantum simulation beyond current methods, by incorporating such phenomena.

\subsection{Reservoir Models}

\begin{figure}
\centering
\includegraphics[width=\linewidth]{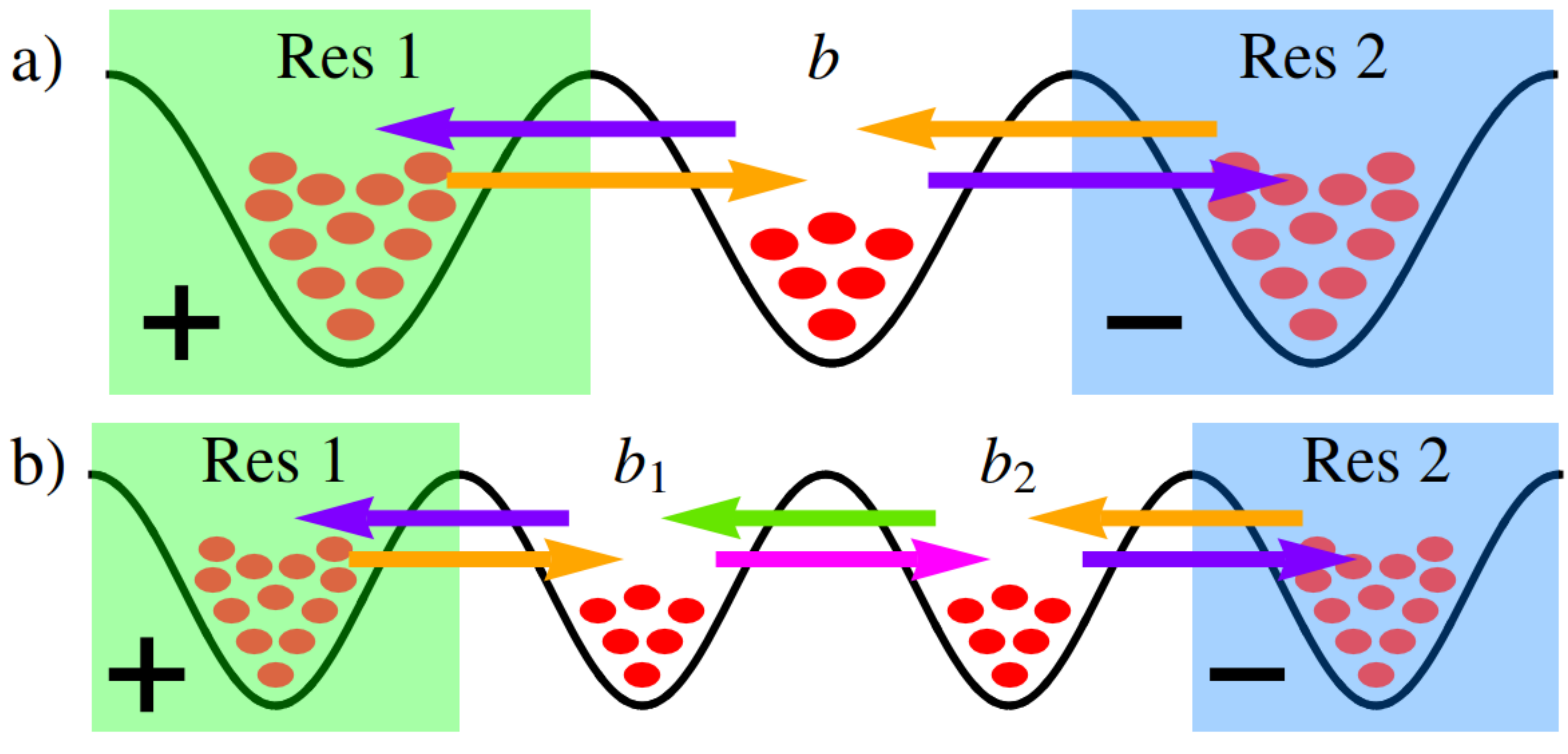}
\caption{Dynamical constraints imposed by measurement of the light allows for selective engineering of atomic dynamics for use in quantum simulation. By designating some sites as reservoir modes, (a) effective pair processes and (b) generalised Dicke models can be realised. The highlighting of sites indicates the contribution of their occupation to the measurement value (green positive, blue negative), which is fixed by the quantum Zeno effect. In both cases, the measurement operator is given by $D_{\Pi0}=N_{\mathrm{Res}1}-N_{\mathrm{Res}2}$). Tunnelling events constrained to only occur simultaneously because of the constraints are indicated by identically coloured arrows.}
\label{figreservoir}
\end{figure}

The matter mode structure defined by the light geometry allows the lattice to be partitioned into sets of sites corresponding to each mode. We can assign a subset of these modes as reservoirs, and investigate the dynamics of the remaining sites, subject to the presence of the reservoirs. Consider the conceptual three-site model shown in \figref{figreservoir}(a). In this scenario, a cavity is used to drive dynamics that generate (homogenous amplitude) two-body correlated tunnelling between the sites, as per Eq.~\eqref{eqtwobodyproc} with a single cavity $c$ and uniform light-matter interaction coefficients $J_{\mathrm{nn}}^{c0}$, such that the system is described by 
\begin{align}
H&=H_M+\Omega_{00}|\alpha_0|^2(D_{00}+B_{00})\nonumber\\&+\frac{\Delta_c|C_c|^2}{2}(B_{c0}^\dagger B_{c0}+B_{c0}B_{c0}^\dagger).
\end{align}
Neglecting the effective (uniform) chemical potential due to the pump light (which can be tuned away by e.g.~using another classical light source), and the on-site interactions (which can be tuned away by Feshbach resonances \cite{lewenstein2012}), this becomes
\begin{align}
\label{eq3site}
H&=\sum_{<ij>}(J^T_{\mathrm{nn}}+\Omega_{00}|\alpha_0|^2J^{00}_{\mathrm{nn}})b_i^\dagger b_j\nonumber\\&+\Delta_c|C_cJ_{\mathrm{nn}}|^2\sum_{\substack{<ij>\\<kl>}} b_i^\dagger b_j b_k^\dagger b_l.
\end{align}

We label the sites $i=1,2,3$, and designate the outer two sites as the reservoirs. The measurement cavity is arranged at the maxima of two coherent antiphase pump lasers antisymmetric about the central site, such that they fully destructively interfere at this site, and have opposing contributions at the corresponding site pairs about the centre. The resulting total pump mode function is $u_0(\bm{r})=u_p(\bm{r})-u_p(-\bm{r})$, and thus the measured operator is $D_{\Pi0}=N_1-N_3$. An example set of appropriate pump modefunctions are $u_p(\bm{r})=\mathrm{sin}(\sqrt{2}\pi x/d)$, where $d$ is the lattice spacing and $x$ is the lattice axis, such modefunctions being achievable by using travelling waves angled at $45^\circ$ to the lattice (the incommensurate nature of the pump with the lattice ensures that any other sites adjacent to the reservoirs do not have vanishing contributions to the measurement, preventing events where one atom tunnels into the central site simultaneously with an atom tunnelling to an external site).  The measurement then constrains the system to remain in a subspace of states in which this is constant. When the measured value for this operator is $\Delta N$, the appropriate projectors applied to the system are (using the designation $\ket{N_1,N_2,N_3}$)
\begin{equation}
\label{eqproj3}
\mathcal{P}_{\Delta N}=\sum_{N_2,N_3}\ket{N_3+\Delta N,N_2,N_3}\bra{N_3+\Delta N, N_2, N_3}.
\end{equation}

Applying these projectors to the Hamiltonian Eq.~\eqref{eq3site}, we obtain the Zeno Hamiltonian $\mathcal{P}_{\Delta N}H\mathcal{P}_{\Delta N}$. The projectors eliminate all of the single-atom tunnelling terms, and preserve only the two-body tunnellings in which $N_1-N_3$ is conserved. These are the following terms: $b_1^\dagger b_2 b_2^\dagger b_1$; $b_2^\dagger b_1 b_1^\dagger b_2$; $b_2^\dagger b_3 b_3^\dagger b_2$; $b_3^\dagger b_2 b_2^\dagger b_3$; $b_1^\dagger b_2 b_3^\dagger b_2$; $b_3^\dagger b_2 b_1^\dagger b_2$; $b_2^\dagger b_1 b_2^\dagger b_3$; and $b_2^\dagger b_3 b_2^\dagger b_1$. The first four of these terms correspond to processes where an atom leaves one of the reservoirs, concurrent with another atom entering the same reservoir, while the latter four describe events where atoms simultaneously leave (or enter) both reservoirs. 

Consider now the reservoirs to both be prepared in coherent states (as would be approximately expected for a system initially in a superfluid state from the Gutzwiller ansatz \cite{lewenstein2012}). Replacing the operators for the two reservoir sites by their coherent amplitudes $b_1\to\beta_1$ and $b_3\to\beta_2$, we now only have one dynamical variable, describing the occupation of the central site. Relabelling this as $b_2\to b$, and defining $n=b^\dagger b$, the effective Hamiltonian becomes
\begin{equation}
H=\Delta_c|C_cJ^{c0}_{\mathrm{nn}}|^2((|\beta_1|^2\!+|\beta_2|^2)(2n+1)+(2\beta_1\beta_2b^\dagger b^\dagger\! +h.c.)).
\end{equation}
Discarding the constant term, and further tuning on-site terms to eliminate the effective chemical potential, we arrive at a Hamiltonian that describes effective pair creation and annihilation dynamics at the central site:
\begin{equation}
H_{\mathrm{PP}}=(\lambda b^\dagger b^\dagger + h.c.),
\end{equation}
where
\begin{equation}
\label{eqlambdares3}
\lambda=2\beta_1\beta_2\Delta_c|C_cJ^{c0}_{\mathrm{nn}}|^2.
\end{equation}
With the experimental parameters considered above, and for typical occupations $|\beta|^2\sim\mathcal{O}(1-10)$, these processes can be of a comparable size to the classical tunnelling rate $J^T_{\mathrm{nn}}$ (or even slightly larger depending on $|\beta|^2$).

Now consider the extension of this to include an additional site between the reservoirs, as depicted in \figref{figreservoir}(b). Again, the outer sites are designated as the reservoirs, and their occupation number difference is the subject of measurement. One way to achieve such a mode structure is to again use the interference of  two coherent pump beams, now arranged to have vanishing contributions on the central two sites. Thus, the measured operator is (labelling the central two sites 1 and 2, and the reservoirs Res1 and Res2) given by $D_{\Pi0}=N_{\mathrm{Res1}}-N_{\mathrm{Res2}}$, again fixed at some value $D_{\Pi0}=\Delta N$, now with the associated projectors
\begin{align}
\mathcal{P}_{\Delta N}=\sum_{N_1,N_2,N_{\mathrm{Res}}}&\ket{N_{\mathrm{Res}}+\Delta N,N_1,N_2,N_{\mathrm{Res}}}\nonumber\\ \times&\bra{N_{\mathrm{Res}}+\Delta N, N_1, N_2,N_{\mathrm{Res}}}.
\end{align}

As before, this measurement imposes constraints on the allowed tunnelling events. Unlike the previous case however, some single atom tunnelling events survive: those between the two central sites $b_1^\dagger b_2$ and $b_2^\dagger b_1$. The permitted correlated tunnelling events involving the reservoirs are analogous to the previous case, with the processes involving atoms simultaneously crossing the same boundary between a reservoir and central site in opposite directions, or the simultaneous tunnelling of a particle from (to) each of the reservoirs in to (from) each of the central sites.

However, there are also now present correlated events where both tunnelling events take place between the central two sites, these being of the form $b^\dagger_1b_2b^\dagger_1b_2$, $b^\dagger_1b_2b^\dagger_2b_1$, $b^\dagger_2b_1b^\dagger_1b_2$, and $b^\dagger_2b_1b^\dagger_2b_1$. Once again replacing the operators for the reservoirs by their coherent state amplitudes, we can write the Hamiltonian describing the dynamics of the central two sites (again for now neglecting the effective chemical potentials) as
\begin{align}
\label{eqfulldicke}
H&=(J^T_{\mathrm{nn}}+\Omega_{00}|\alpha_0|^2J^{00}_{\mathrm{nn}})(b_1^\dagger b_2+h.c.)\nonumber\\
&+(2\beta_1\beta_2\Delta_c|C_cJ^{c0}_{\mathrm{nn}}|^2b_1^\dagger b_2^\dagger+h.c.)\nonumber\\
&+\Delta_c|C_cJ^{c0}_{\mathrm{nn}}|^2(b^\dagger_1b_2b^\dagger_1b_2\!+\!b^\dagger_1b_2b^\dagger_2b_1\!+\!b^\dagger_2b_1b^\dagger_1b_2\!+\!b^\dagger_2b_1b^\dagger_2b_1).
\end{align}

Consider now the reservoirs to have a large occupation compared to the central sites (e.g.~$\mathcal{O}(1)$ atom per central site, and $\mathcal{O}(10)$ per reservoir; such filling factors are available in typical experiments \cite{klinder2015,landig2016}). In this case, the terms in the third line of Eq.~\eqref{eqfulldicke}, corresponding to the correlated processes between atoms in the central sites alone, become negligible compared to the reservoir-based correlated tunnellings. Reintroducing the chemical potential (which, as noted before, can be tuned with additional semiclassical light sources), the effective Hamiltonian describing the central two sites is now a generalised Dicke Hamiltonian \cite{dicke1954, schmidt2014}:
\begin{equation}
\label{eqgdicke}
H_{\mathrm{GD}}=\sum_{i=1}^2 \mu_i b^\dagger_i b_i + (\lambda_1 b^\dagger_1b_2 + \lambda_2 b_1^\dagger b_2^\dagger +h.c.),
\end{equation}
where
\begin{align}
\lambda_1&=J^T_{\mathrm{nn}}+\Omega_{00}|\alpha_0|^2J_{\mathrm{nn}}^{00}\\
\lambda_2&=2\beta_1\beta_2\Delta_c|C_cJ^{c0}_{\mathrm{nn}}|^2.
\end{align}

In contrast to the original model, and its corresponding realisation in optical cavities \cite{baumann2010}, the parameters $\lambda_{1,2}$, which represent the co- and counter-rotating terms respectively, can be tuned independently, by adjusting the optical geometry; their magnitude is controlled by e.g.~increasing the pump strength or adjusting the reservoir populations $\beta_{1,2}$, and complex phases can be induced with the use of additional pump beams. 

\begin{figure}
\centering
\includegraphics[width=\linewidth]{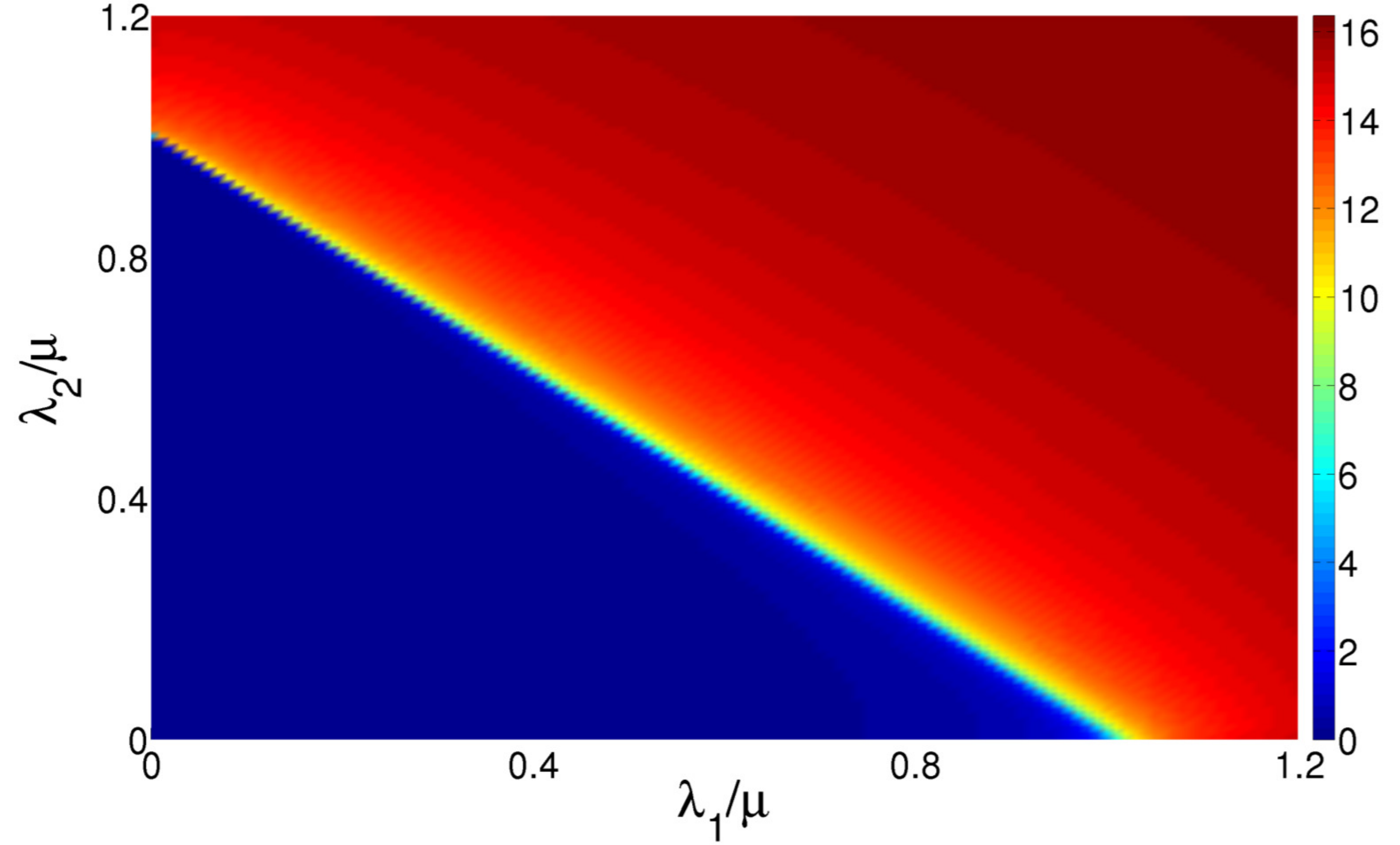}
\caption{Ground state average occupation $\langle n_1 \rangle$ of the generalised Dicke model, showing a change in the superradiance phase transition point when varying $\lambda_1/\lambda_2$. In the standard Dicke model, the transition occurs at $\lambda_1=\lambda_2=\mu/2$; here we find the general boundary at $\lambda_1+\lambda_2\approx\mu$. We use $\mu_1=\mu_2=\mu$, and a maximum occupation of 20 for each mode.}
\label{figphase}
\end{figure}

In addition to the well-known phase diagrams for $\lambda_1=\lambda_2$ (traditional Dicke) and $\lambda_2=0$ (Tavis-Cummings) \cite{bastarrachea2014} in the quantum case, classical treatments have shown bifurcations when varying $\lambda_{1,2}$ \cite{aguiar1991}, suggesting possible further novel phase behavior. We explore part of this extended parameter space by using exact diagonalisation methods to find the ground state average occupation of each of the modes. Specifically, we do this by limiting the occupation of the two modes to 20 atoms per site, and with this limitation we can construct the Hamiltonian Eq.~\eqref{eqgdicke}, and obtain the ground state. We do this for the regime in which $\mu_1=\mu_2=\mu$, and find that a general phase boundary [\figref{figphase}] for the onset of superradiance (that is, the transition from $\langle n_j \rangle=0$ to diverging (in the full non-occupation-limited case)) occurs at $\lambda_1+\lambda_2\approx\mu$, with the standard transition at $\lambda_1=\lambda_2=\mu/2$ being a special case of this.

This can be further extended by using the light measurement to fix occupation number differences between larger numbers of modes \cite{elliott2015}, allowing for additional reservoir modes to be generated.  These can be used to increase the number of sites coupled to reservoirs, and increase the number of simulated modes in our model (for example, to have multiple atomic species). As a concrete example of how these building blocks we propose can be put together to form more complex systems, we now describe how to realise a generalised Dicke model with two (synthetic) atomic species and one synthetic light mode. 

\begin{figure}
\centering
\includegraphics[width=\linewidth]{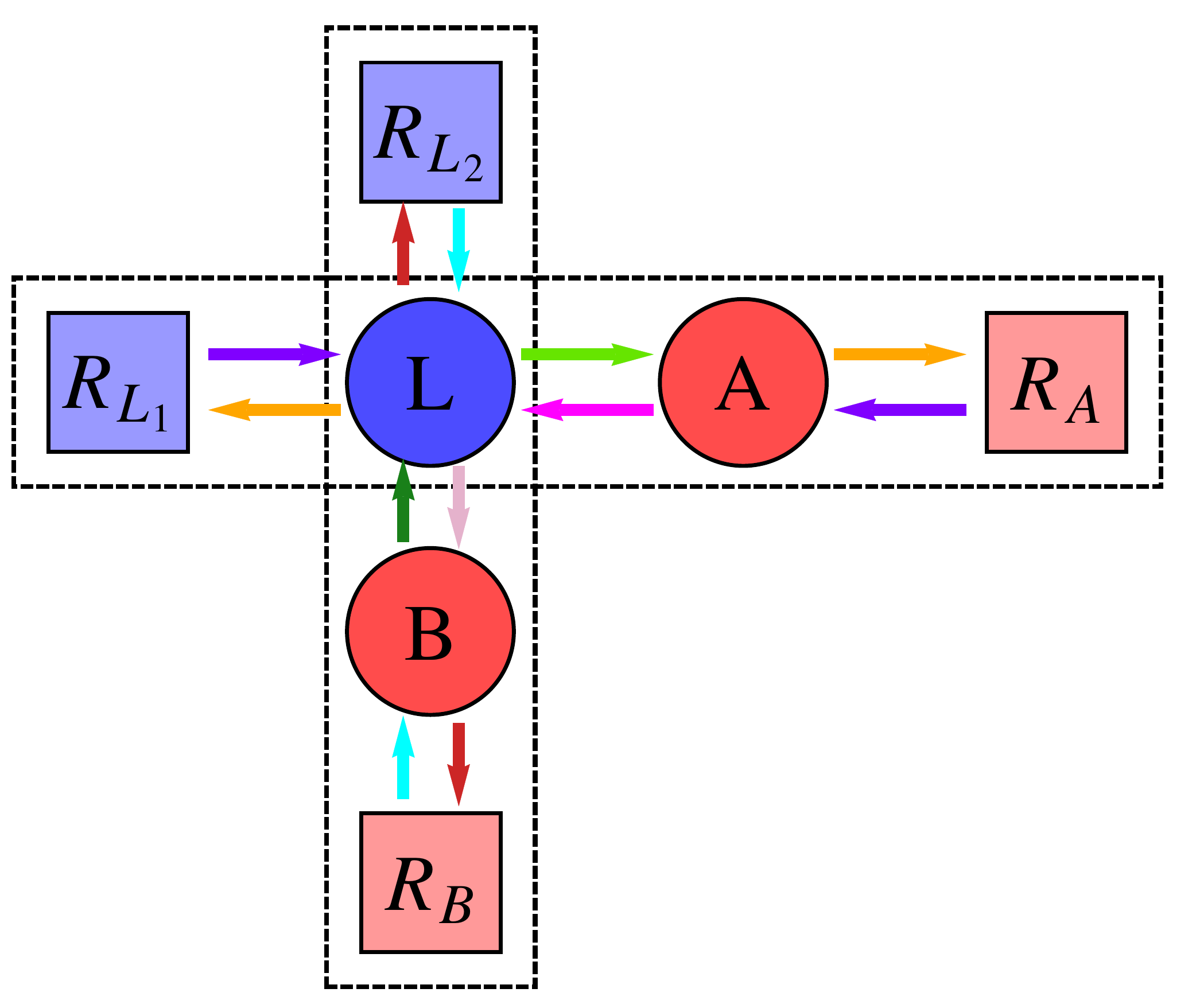}
\caption{Setup for the two-species generalised Dicke model. Combining two setups for the one-species generalised Dicke model, with a mutual synthetic light mode, allows the introduction of another atomic species. Circles represent the simulated modes, squares the reservoirs. The required measurements made are of occupation differences $R_{L_1}-R_{A}$ and $R_{L_2}-R_{B}$, which correlate certain tunnelling events, as indicated by arrows the same colour.}
\label{figtwodicke}
\end{figure}

Consider an amalgamation of two copies of the above setup for the one-species generalised Dicke model. When these setups are crossed perpendicularly, with both setups sharing a common site for the synthetic light mode (see \figref{figtwodicke}), the two-species generalised Dicke model can be realised. As before, cavities are used to induce the correlated tunnelling dynamics, one in each of the setups (that is, the correlated events both involve tunnellings within the same individual setup). By using also two measurement cavities, to fix each the difference in occupation of the reservoir sites in one of the setups (to $\Delta N_1$ and $\Delta N_2$ respectively), the resulting projectors are
\begin{align}
\mathcal{P}_{\Delta N_1\Delta N_2}&=\!\!\!\!\!\!\!\sum_{\substack{N_L,N_A,N_B,\\N_{R1},N_{R2}}}\!\!\!\!\!\!\!\!\ket{N_{R1}+\Delta N_1,N_{R2}+\Delta N_2,N_{R1},N_{R2}}\nonumber\\ \times&\bra{N_{R1}+\Delta N_1,N_{R2}+\Delta N_2,N_{R1},N_{R2}}\nonumber\\\otimes&\ket{N_A,N_B,N_L}\bra{N_A,N_B,N_L},
\end{align}
where $A$ and $B$ are the simulated atomic species, $L$ is the synthetic light mode, $R_{L_1}$ and $R_{A}$ are the reservoir modes associated with the simulated light and atomic modes respectively in the first setup (and $R_{L_2}$ and $R_{B}$ for the second), and the states are labelled as $\ket{N_{R_{L_1}},N_{R_{L_2}},N_{R_A},N_{R_B}}\otimes\ket{N_A,N_B,N_L}$.

Following the analysis for the single-species Dicke model Eq.~\eqref{eqgdicke}, replacing the reservoirs by coherent states, and noting that both setups have a central site in common (corresponding to the synthetic light mode), the resulting Hamiltonian in the Zeno subspace for the central sites may be written
\begin{equation}
H_{\mathrm{2GD}}=\!\!\!\!\sum_{i=A,B,L}\!\!\!\! \mu_i b^\dagger_i b_i  + \!\sum_{i=A,B}\!(\lambda_1^{(i)} b^\dagger_ib_L + \lambda_2^{(i)} b_i^\dagger b_L^\dagger +h.c.),
\end{equation}
where the $\lambda_{1,2}^{(i)}$ are the same as for the one-species setup, with the appropriate coefficients  $\beta_{1,2}$, $\Delta_c$, $C_c$ and $J_{\mathrm{nn}}^{c0}$ as for each of the individual setups.

Yet more exotic setups can be envisaged within this framework, through the use of additional reservoirs, or with multiple sites coupling to each reservoir. In the extreme case of each site in the simulated system being coupled to its own two reservoirs with fixed occupation differences (i.e.~a lattice of copies of the setup \figref{figreservoir}(a)), one would obtain an extended Bose-Hubbard model with additional pair creation/annihilation effects present at each site. Perhaps more straightforward to realise experimentally, consider a similar setup with two reservoir modes (again with fixed occupation difference) common to all sites. This could be achieved by again taking the setup of \figref{figreservoir}(a), where instead of three individual sites we instead have three connected lattices (which may be one- or two-dimensional), with each of the lattices forming one of the three modes. Arranging the measurement cavity to scatter light with the same phase from all sites in a given reservoir mode renders them indistinguishable to the measurement, as per the matter mode structure, and so they behave like a collective reservoir. In the superfluid regime, each site in reservoir $j$ may be approximated again by the Gutzwiller ansatz, and described by a coherent state of amplitude $\beta_j$, where $|\beta_j|^2$ is the filling factor of the lattice. As the sites in each reservoir mode are indistinguishable, the correlated tunnellings from each of the reservoirs to/from the central lattice may occur due to any of the sites in each reservoir. Thus, the effective pair creation/annihilation in the central lattice may have the pair of atoms far apart in the lattice from each other. We have analogous projectors to the three-site case Eq.~\eqref{eqproj3}, but with each number now representing the occupation across the whole of the respective lattice, rather than individual sites. Note that the standard Bose-Hubbard dynamics for atoms tunnelling between sites in the central lattice is unaffected by the measurement, and thus the resulting system in the central lattice obeys a Bose-Hubbard model with long-range pair creation/annihilation;
\begin{align}
H_{\mathrm{PPBHM}}&=-J\realsum_{<ij>}b^\dagger_ib_j + \frac{U}{2}\realsum_i b^\dagger_ib^\dagger_ib_ib_i\nonumber\\&+(\lambda\realsum_{ij}b^\dagger_ib^\dagger_j+h.c.),
\end{align}
where
\begin{equation}
\lambda=2\beta_1\beta_2\Delta_c|C_cJ^{c0}_{\mathrm{nn}}|^2.
\end{equation}

\subsection{Dynamical Global Gauge Fields}

Another application of this framework for quantum simulation is in the synthesis of artificial dynamical gauge fields. Current proposals to this end also typically employ ultracold atoms in optical lattices, focussing on local gauge fields based on quantum link models \cite{horn1981, orland1990, brower1999}, where changes in the gauge field due to the motion of matter are simulated by the motion of another particle species that plays the role of the gauge field \cite{banerjee2012, banerjee2013, stannigel2014}. In addition to the possibility of extending the variety of such models by the inclusion of long-range interactions, the introduction of the cavity-mediated dynamics presents a further opportunity: the long-range nature of the interactions allows for the correlated atomic motion itself to be long-range, and hence the links need not be local. As such, this paves the way for realising global dynamical gauge fields, where motion across all sites is controlled by a common link.

For example, consider a one-dimensional lattice, with light measurement of a site-dependent strength. Specifically (by, e.g.~a gradiated intensity of the measurement pump), the measured light state has an amplitude that is proportional to $D_{\Pi0}=\sum_jj\Upsilon N_j$, for some constant $\Upsilon$. We consider two auxiliary sites $L$ and $R$, between which atoms can tunnel (and tunnelling out of this pair is suppressed), which are also measured in the same fashion, contributing an effective $\Upsilon N_R$ (plus constant) to the measurement value (because their conserved total occupation $N_L+N_R$). In this case, sites $L$ and $R$ form a link that mediates the dynamics in the rest of the lattice, as the Zeno Hamiltonian will then only contain the dynamics from the cavity-mediated correlated tunnelling events, where an atom tunnels in the main lattice simultaneously with a tunnelling event either in the link, or in the opposite direction in the main lattice. This follows from considering that a tunnelling $j\to j+1$ (or $L\to R$) increases the value of $D_{\Pi0}$ by $\Upsilon$, while a tunnelling $j\to j-1$ (or $R\to L$) decreases it by $\Upsilon$, and so only when one of each of these processes occur simultaneously is the measurement outcome value preserved. 

The resulting Hamiltonian after applying this constraint is 
\begin{align}
H&=\lambda\sum_j (b^\dagger_jb_{j+1}(\sum_kb^\dagger_{k+1}b_{k} + \vartheta b_L^\dagger b_R)+h.c.)\nonumber\\&+\lambda\vartheta^2(b^\dagger_Lb_Rb^\dagger_Rb_L+h.c.).
\end{align}
where 
\begin{equation}
\lambda=\Delta_c|C_cJ_{\mathrm{nn}}^{c0}|^2
\end{equation}
and $\vartheta^2$ is the ratio of the intensity of the pump used to drive dynamics at the link sites compared to the rest of the lattice. Equivalently, by mapping the link to a spin with $2S^Z=N_L-N_R$, this can be written
\begin{align}
H_{\mathrm{DGGF}}&=\lambda \sum_j(b^\dagger_jb_{j+1}(\sum_kb^\dagger_{k+1}b_{k} + \vartheta S^+) + h.c.)\nonumber\\& - 2\lambda \vartheta^2 (S^Z)^2.
\end{align}
This $\lambda$ is comparable in size to the equivalent parameter considered in the reservoir models (e.g. Eq.\eqref{eqlambdares3}), and so too may also be of a similar magnitude to the standard tunnelling $J^T_{\mathrm{nn}}$, with its precise value depending on the pump strength.

This forms a global-link dynamical matter-gauge field interaction, in contrast to current proposals for dynamical gauge fields, which are limited to local links by their finite-range interactions. Strictly, we note that pair-correlated tunnelling events are allowed within the main lattice independent of the link, provided that they occur in opposite directions, though the particle current across the lattice is wholly dependent on the link. These terms bypassing the link can be made less significant by adjustment of $\vartheta$, which also controls the gauge field energy terms $\propto (S^Z)^2$. These gauge field energy terms can be further engineered through the density-density interactions discussed above. Unlike the local link models ubiquitous in high-energy physics, the common global link here leads to a peculiar effect where the motion of a particle at any site can significantly affect the field experienced by all other particles, at all other sites.

\section{Conclusions}

In summary, we have characterised the new dynamics manifest by the interactions of atomic quantum gases with quantum light in optical cavities, exhibiting effects beyond those possible with classical light, and subsequently shown that these may then be controlled through measurement of the light leaked from the cavity. These effects include long-range correlated tunnelling, effective pair processes, and density-density interactions. Further, we have discussed how this provides opportunities for the enhancement of quantum simulations, by using these correlated processes. Specifically, we have demonstrated how the formalism can mimic superexchange interactions, reservoir models, and dynamical global gauge fields. This invites a wealth of opportunities for further study, such as combining the various simulation building blocks presented here to generate yet more exotic and interesting systems for study, finding additional building blocks to expand the simulation framework, and further characterisation of cavity-induced processes.

As discussed above, our proposal should be feasible with current state-of-the-art experimental setups. So far, several groups have trapped Bose-Einstein condensates inside cavities, without a lattice potential \cite{baumann2010,wolke2012,schmidt2014}, while others have scattered light from quantum gases trapped in lattice potentials, but with no cavity present \cite{weitenberg2011,miyake2011}. The amalgamation of these proposals has recently been achieved, where the light scattered into an optical cavity by the atoms generates a further, quantum potential for the atoms, which is dynamically evolving conditional on the atomic state \cite{klinder2015, landig2016}. These experiments already exhibit a particular case of the cavity-induced dynamics, where the the cavity field causes the atoms to self-organise into charge density wave and supersolid phases.

Additionally, many recent experiments have demonstrated examples of quantum Zeno physics in similar, but less versatile settings, such as cavity QED and optical lattices \cite{patil2014, schafer2014, signoles2014, barontini2015}. Thus, the possibility to use measurement for such a selective suppression of dynamics is well verified. Furthermore, it may be possible to implement the dynamical effects discussed here using other types of systems that are also based on off-resonant scattering, such as molecules \cite{mekhov2013}, fermions \cite{ruostekoski2009}, spins \cite{cordobes2014}, ions \cite{blatt2012}, and semiconductor \cite{trauzettel2007} and superconducting qubits \cite{fink2009}, as their dynamics are based on similar mathematical structures.

\section*{Acknowledgements}
The authors thank the Engineering and Physical Sciences Research Council for financial support (Doctoral Training Account and EP/I004394/1).

\appendix
\section*{Effective Heisenberg Equations of Atomic Operators}
\label{secheisenberg}
We here provide some more elaboration on the need for symmetrisation in the effective atomic Hamiltonian Eq.~\eqref{eqheff} that we mentioned in Section \ref{secmodel}. First note that before the adiabatic elimination, the light-matter interaction Hamiltonian Eq.~\eqref{eqHlm} contains terms proportional to $a_m^\dagger a_n b^\dagger_ib_j$, which may be re-ordered as $b_i^\dagger b_j a_m^\dagger a_n$ with impunity, as the light and matter operators commute. This is no longer true once the light modes are replaced by their steady state operators, which are now functions of the matter operators through the $\mathcal{J}_{mn}$. Explicitly, let us consider the time evolution of the matter operators due to the $H_{LM}$ prior to the adiabatic elimination (we may here ignore the contribution from the bare $H_M$ as it is unchanged by the elimination). This is given by
\begin{align}
\dot{b}_k&=i[H_{LM},b_k] \nonumber \\
&=-i\sum_{mn}\Omega_{mn}a^\dagger_ma_n\sum_{j}J_{kj}^{mn}b_j.
\end{align}
After replacing the light operators by their steady states, this will read (neglecting the cavity-cavity terms)
\begin{align}
\label{equnsym}
\dot{b}_k&=-i\Omega_{00}|\alpha_0|^2\sum_jJ^{00}_{kj}b_j \nonumber\\
&+\sum_{m}\Omega_{m0}\alpha_0^*C_m\sum_{ijl}J^{m0}_{ij}{J^{m0}_{kl}}^*b_i^\dagger b_j b_l \nonumber \\
&+\sum_{m}\Omega_{0m}\alpha_0C_m^*\sum_{ijl}{J^{m0}_{ij}}^*J^{m0}_{kl}b_i^\dagger b_j b_l 
\end{align}

However, the replacement of the light operators by the steady state in the Hamiltonian will result in terms of the form $b_i^\dagger b_j b^\dagger_{i'}b^\dagger_{j'}$, which give rise to terms in the Heisenberg equation of the form $b_i^\dagger b_j b_l$ (present in Eq.~\eqref{equnsym}), but also $b_l b_i^\dagger b_j$ (which are not). If one symmetrises the light and matter operators before the elimination, then we instead have 
\begin{align}
H_{LM}=\frac{1}{2}\sum_{mn}\Omega_{mn}(&a^\dagger_ma_n\sum_{ij}J_{ij}^{mn}b_i^\dagger b_j\nonumber \\+&\sum_{ij}J_{ij}^{mn}b_i^\dagger b_ja^\dagger_ma_n),
\end{align}
which has a corresponding evolution equation
\begin{align}
\dot{b}_k&=\frac{-i}{2}(\sum_{mn}\Omega_{mn}a^\dagger_ma_n\sum_{j}J_{kj}^{mn}b_j \nonumber \\
&+\sum_{mn}\Omega_{mn}\sum_{j}J_{kj}^{mn}b_ja^\dagger_ma_n).
\end{align}
After replacement of the light operators, this now contains the required $b_lb^\dagger_ib_j$ terms, and corresponds to the same evolution equation as that yielded by the effective Hamiltonian after adiabatic elimination Eq.~\eqref{eqheff}. Further discussion on this aspect of adiabatic elimination of the light modes may be found in \cite{maschler2008}.

\bibliography{ref}

\end{document}